\documentclass[conference, a4paper]{IEEEtran}
\IEEEoverridecommandlockouts
\usepackage{cite}
\usepackage{amsmath,amssymb,amsfonts}
\usepackage{algorithmic}
\usepackage{graphicx}
\usepackage{textcomp}
\usepackage{xcolor}
\usepackage{makecell} 
\usepackage{algorithm}
\usepackage{algorithmic}
\usepackage{array}
\usepackage{graphicx} 
\usepackage{subcaption} 
\usepackage{booktabs}
\usepackage{diagbox} 
\usepackage{array}    
{\footnotesize }
\def\BibTeX{{\rm B\kern-.05em{\sc i\kern-.025em b}\kern-.08em
    T\kern-.1667em\lower.7ex\hbox{E}\kern-.125emX}}
\begin{document}

\title{Perception-Enhanced Multitask Multimodal Semantic Communication for UAV-Assisted Integrated Sensing and Communication System
{\footnotesize }

}

\author{\IEEEauthorblockN{Ziji Guo\textsuperscript{\dag}, Haonan Tong\textsuperscript{\dag},
Zhilong Zhang\textsuperscript{\dag}, Danpu Liu\textsuperscript{\dag}}
\IEEEauthorblockA{{\dag}Beijing Key Laboratory of Network System Architecture and Convergence, Beijing Advanced Information Network Laboratory,\\
Beijing University of Posts and Telecommunications, Beijing, China 100876.}
\text{Email:} zijiguo@bupt.edu.cn, hntong@bupt.edu.cn, zhangzhilong@bupt.edu.cn, dpliu@bupt.edu.cn
}
\vspace{-0.4cm}
\maketitle
\vspace{-2cm}
\begin{abstract}
Recent advances in integrated sensing and communication (ISAC) unmanned aerial vehicles (UAVs) have enabled their widespread deployment in critical applications such as emergency management. This paper investigates the challenge of efficient multitask multimodal data communication in UAV-assisted ISAC systems, in the considered system model, hyperspectral (HSI) and LiDAR data are collected by UAV-mounted sensors for both target classification and data reconstruction at the terrestrial BS. The limited channel capacity and complex environmental conditions pose significant challenges to effective air-to-ground communication. To tackle this issue, we propose a perception-enhanced multitask multimodal semantic communication (PE-MMSC) system that strategically leverages the onboard computational and sensing capabilities of UAVs. In particular, we first propose a robust multimodal feature fusion method that adaptively combines HSI and LiDAR semantics while considering channel noise and task requirements. Then the method introduces a perception-enhanced (PE) module incorporating attention mechanisms to perform coarse classification on UAV side, thereby optimizing the attention-based multimodal fusion and transmission. Experimental results demonstrate that the proposed PE-MMSC system achieves 5\%--10\% higher target classification accuracy compared to conventional systems without PE module, while maintaining comparable data reconstruction quality with acceptable computational overheads.
\end{abstract}

\addtolength{\rightmargin}{0.095in} 
\section{Introduction}
The emergence of the low-altitude economy has significantly advanced the research of unmanned aerial vehicles (UAVs) in industrial applications\cite{16}, particularly in mission critical domains such as disaster response \cite{1}. Modern UAV with advantages of rapid deployment capability, adaptive mission scheduling flexibility, and aerial line-of-sight (LoS) coverage that attributes not only satisfy fundamental communication demands but also enable UAVs to excel in remote sensing operations\cite{2}, including precision disaster assessment and dynamic target localization. The increasing complexity of aerial missions now demands concurrent execution of heterogeneous tasks, particularly sensing and communication functions, under constrained computational resources. This operational imperative has driven the development of integrated sensing and communication (ISAC) architectures, which enable unified coordination between multiple tasks \cite{3}. 

In recent years, ISAC has been extended from traditional millimeter wave radar-based perception to multimodal sensor integration on UAVs, which improves communication performance through sensor collaborations \cite{4}. This evolution has driven growing interest in multimodal fusion within remote sensing \cite{6}. Fusion methods were classified into symmetric approaches (including fusion at the data level, semantic level, and decision level) and asymmetric approaches based on the data hierarchy in \cite{7}. However, the piror arts predominantly treat multimodal data as simple multi-channel inputs, to address these limitations, the scale inconsistency was resolved in \cite{8} across modalities through pyramid structures and multi-step fusion networks, while \cite{9} adapted multimodal channel weights according to Line-of-Sight (LoS) conditions. Besides, in many scenarios, it is not sufficient to merely accomplish a single task, rather, multiple tasks need to be completed using remote sensing data. EndNet, proposed in \cite{13}, addressed the requirements of multimodal fusion for  target classification and data reconstruction. Meanwhile, Image Transmission and Performance Analysis (MTP) in \cite{14} considered the needs of segmentations, image classification, and change detection, as discussed. Unfortunately, existing solutions introduce high computational costs and assume data acquisition and task execution on the same side, neglecting bandwidth constraints and environmental noise impacts on transmission.

The bandwidth limitations and dynamic characteristics of UAV communications have promoted semantic communication as an effective solution \cite{18}\cite{s3}. In \cite{17}, it was shown that semantic communication can substantially reduce data transmission volumes while maintaining communication robustness under low SNR conditions. To enhance semantic communication reliability in remote sensing field, a guide-inspired Transformer block was designed in \cite{10} to build decoders that improve the accuracy of semantic information extraction. Furthermore, a multipath atrous module was proposed in \cite{11} to address progressive semantic segmentation tasks, further advancing the efficiency of semantic communication in complex environments. Moreover, multisource multimodal semantic communication has been widely discussed. In \cite{s2}, a diffusion - model (DM) - based channel enhancer (DMCE) is proposed to extract and compress multi - source data features. In \cite{s1}, an efficient multimodal data communication scheme for video conferencing is studied, which can effectively compress data while ensuring video transmission quality. Nevertheless, current approaches predominantly utilize raw remote sensing data as what to be transmitted and fused, lacking explicit task-oriented designs and in-depth analysis in multimodal fusion processes.

A perception-enhanced multitask multimodal semantic communication framework for UAV-ISAC networks is studied in this paper. We consider UAVs equipped with multimodal sensors capable of collecting complementary hyperspectral (HSI) and LiDAR remote sensing data. These data are transmitted to the base station (BS), which concurrently executes two mission-critical tasks: target classification and data reconstruction. Building upon the synergistic design of ISAC, we propose to exploit the UAV's inherent sensing capabilities to optimize semantic information fusion and transmission efficiency. The principal contributions of this work are as follows:
\begin{itemize}
    \item[\textbullet] A perception-enhanced multitask multimodal semantic communication (PE-MMSC) framework for UAV-assisted ISAC is proposed. In multitask scenarios, we consider the limited computational power of UAVs, and introduce a Perception Enhancement (PE) module on the UAV side, which is a lightweight neural network, to enhance the performance of multitask.
    \item[\textbullet] We design a specific PE module with an attention mechanism. The PE module performs preliminary coarse classification on the UAV side to provide an interference-free classification result as a reference, thereby adjusting the channel attention in the multimodal fusion process. 
    \item[\textbullet] The proposed framework is tested on a real-world remote sensing dataset. The results demonstrate that the PE-MMSC system achieves significant performance improvements with minimal computational cost. The target classification accuracy is improved by 5\% to 10\% while ensuring data reconstruction quality compared to traditional system without PE module.
\end{itemize}

\section{System Model}
\label{sec:sys}
\begin{figure*}[ht] 
    \centering 
    \includegraphics[width=1\textwidth]{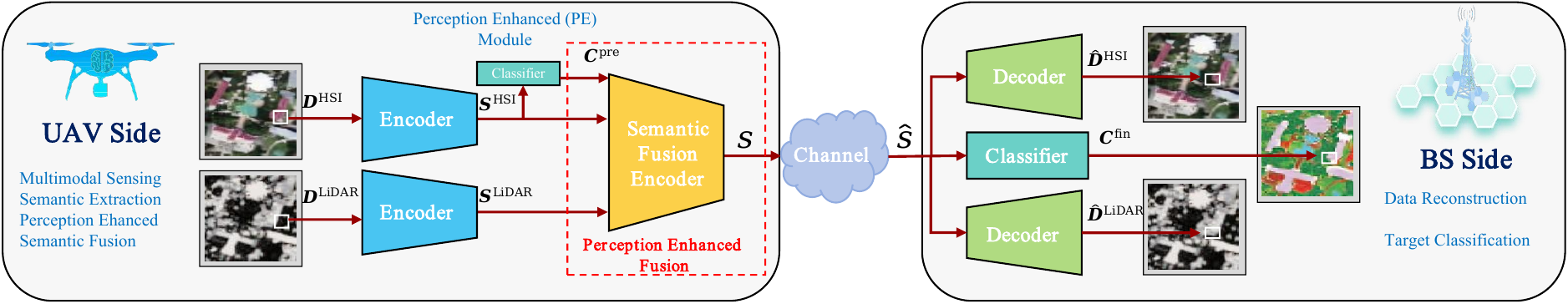} 
    \caption{Perception-enhanced multitask multimodal semantic communication for UAV-assisted ISAC system} 
    \label{fig:sys} 
    \vspace{-10pt}
\end{figure*}
\subsection{System Architecture}

We investigate a scenario where UAV collaborate with BS to accomplish multitask. To achieve comprehensive environmental perception and ensure task quality under low SNR levels, the UAV equipped with multimodal sensors extracts semantic features from the sensing data, fuses them and transmit to the BS. In practical applications, different tasks necessitate diverse requirements, BS is required to perform multitasking with fused data, including target classification and data reconstruction, as illustrated in Fig.\ref{fig:sys}.

Taking into account the distinct emphases that different remote sensing modalities place on characterizing the environment, we chose HSI data $ \boldsymbol{D}^{\text{HSI}} $ and LiDAR data $ \boldsymbol{D}^{\text{LiDAR}} $ as sensing data. $ \boldsymbol{D}^{\text{HSI}} $ contains 2D data from multiple frequency range, while $ \boldsymbol{D}^{\text{LiDAR}} $ provides 3D spatial information. In addition, the semantic extraction and fusion techniques employed reduce information redundancy between modalities, thus decreasing the amount of transmitted data, which is particularly beneficial in UAV communication scenarios constrained by limited bandwidth. 

Semantic communication is used to against the channel noise and fading, ensuring transmission quality even under low SNR levels, which including PE-MMSC framework and PE module. Next we will introduce them.

The proposed PE-MMSC framework addresses the multitask demands of the BS in complex scenarios, while the PE Module on the UAV. This compensates for the lack of effective fusion guidance in traditional multimodal fusion methods at the semantic level, enabling effective adjustment of channel attention during the semantic fusion processes. The PE module on the UAV side performs outputs preliminary coarse classification result $ \boldsymbol{C}^{\text{pre}} $ to guide the multimodal fusion. Due to its immunity to environmental interference, $ \boldsymbol{C}^{\text{pre}} $ offers valuable classification information, enhancing the UAV to refine the effective information under low SNR levels.

\subsection{Semantic Fusion and Communication Model}
Since $ \boldsymbol{D}^{\text{LiDAR}} $ mainly contain spatial information, which performs less relativity with target classification task, here, we take $ \boldsymbol{D}^{\text{HSI}} $ as the primary modality, with $ \boldsymbol{D}^{\text{LiDAR}} $ serving as the auxiliary modality. $ \boldsymbol{D}^{\text{HSI}} $ and $\boldsymbol{D}^{\text{LiDAR}} $ are first processed through an semantic encoder to extract the semantic features, given by
\begin{equation}
\boldsymbol{S}^{\text{mod}}=f_{\varphi^\text{mod}_\text{E}}^{\text{E}}(\boldsymbol{D}^{\text{mod}}),\; \text{mod}\in \text{\{HSI,LiDAR\}}
\end{equation}
where $f_{\varphi^\text{mod}}^{\text{E}}(\cdot)$ represents the encoder function parameterized by $\varphi^\text{mod}_\text{E}$, and $\boldsymbol{S}^\text{mod}=[s_1^\text{mod},...,s_{n^\text{mod}}^\text{mod}]$ represents the extracted semantic features of different modality with $n^\text{mod}$ is the feature number, and $s_{i^\text{mod}}$ for the $i$-th semantic feature of different modality. Furthermore, $ \boldsymbol{S}^{\text{HSI}} $ is employed on the UAV to perform preliminary coarse classification, which is given by
\begin{equation}
\boldsymbol{C}^{\text{pre}}=f_{\psi^\text{pre}}^{\text{C}}(\boldsymbol{S}^{\text{mod}})
\end{equation}
where $\boldsymbol{C}^{\text{pre}}=[p_1^\text{pre},...,p_m^\text{pre}]$ represents the preliminary coarse classification result with $\text{pre}$ means preliminary, $m$ is the number of target categories, which contributes to subsequent multimodal fusion, and $p_i^\text{pre}$ for the classification probability of $i$-th categories. $f_{\psi^\text{pre}}^{\text{C}}$ represents the preliminary coarse classification function parameterized by $\psi^\text{pre}$.  Subsequently, all single-modal semantic features are fused under the guidance of $\boldsymbol{C}^{\text{pre}}$ to obtain multimodal fusion semantic features as
\begin{equation}
\boldsymbol{S}=f_\omega^F(\boldsymbol{S}^{\text{HSI}},\boldsymbol{S}^{\text{LiDAR}},\boldsymbol{C}^{\text{pre}})
\end{equation}
where $f_\omega^F(\cdot)$ represents the fusion function parameterized by $\omega$, and $\boldsymbol{S}$ is the fused semantic features. Then, $\boldsymbol{S}$ is transmitted via a wireless channel as
\begin{equation}
\boldsymbol{\hat{S}}=h\cdot\boldsymbol{S}+\boldsymbol{n}
\end{equation}
where $\boldsymbol{\hat{S}}$ represents the semantic features received by the BS after transmission degradation, $h$ is Rayleigh channel coefficient\cite{15}, and $\boldsymbol{n} \sim \mathcal{N}(0, \sigma^2 \boldsymbol{I})$ denotes Gaussian channel noise with $\sigma^2$ being noise variance and $\boldsymbol{I}$ being identity matrix.

BS utilizes $\boldsymbol{\hat{S}}$ to perform multiple tasks, including data reconstruction and target classification. The data reconstruction is 
\begin{equation}
\boldsymbol{\hat{D}^\text{mod}}=f_{\varphi^\text{mod}_D}^{\text{D}}(\boldsymbol{\hat{S}})
\end{equation}
where $f_{\varphi^\text{mod}_D}^{\text{D}}(\cdot)$ represents the decoder function parameterized by $\varphi^\text{mod}_D$, and $\hat{D}^\text{mod}$ denotes the sensing reconstructed from the corresponding modality. Target classification is
\begin{equation}
\boldsymbol{C^\text{fin}}=f_{\psi^\text{fin}}^{\text{C}}(\boldsymbol{\hat{S}})
\end{equation}
where $f_{\psi^\text{fin}}^{\text{C}}(\cdot)$represents the multimodal target classification function parameterized by $\psi^\text{fin}$, and ${C}^\text{fin}$ denotes the final classification result with $\text{fin}$ means final.

We consider transmitting $N$ samples, getting sensing data $\boldsymbol{D}_N^{\text{mod}}\in\mathbb{R}^{N \times n^{\text{mod}}}$ in UAV, and finally output data reconstruction results $\boldsymbol{\hat{D}}_N^{\text{mod}} \in \mathbb{R}^{N \times n^{\text{mod}}}$ and target classification result $\boldsymbol{C}_N^{\text{phase}} \in \mathbb{R}^{N \times m}$, with $\text{phase} \in \{\text{pre},\text{fin}\}$ to represent the classification phase. For the data reconstruction, we use Normalized Mean Squared Error($\text{NMSE}$) as an evaluation metric, which is given by
\begin{equation}
\text{NMSE}(\boldsymbol{D}_N^{\text{mod}},\boldsymbol{\hat{D}}_N^{\text{mod}}) = \frac{\mathbb{E}\left[\left\| \boldsymbol{D}_N^{\text{mod}} - \boldsymbol{\hat{D}}_N^{\text{mod}} \right\|^2\right]}{\mathbb{E}\left[\left\| \boldsymbol{D}_N^{\text{mod}} \right\|^2\right]}
\end{equation}
where $\mathbb{E}(\cdot)$ represents expected value function, and $\text{A}(\cdot,\cdot)$ is used to evaluate target classification, given by
\begin{equation}
\text{A}(\boldsymbol{C}_N^{\text{fin}},\boldsymbol{C}_N^{\text{true}}) = \mathbb{E}[\mathbb{I}[\text{argmax}(\boldsymbol{C}_N^{\text{fin}})=\text{argmax}(\boldsymbol{C}_N^{\text{true}})]]
\end{equation}
where $\text{argmax}$ means argument of the maximum, $\boldsymbol{C}_N^{\text{true}}$ represents ground truth samples, shaped in one-hot format.
\begin{equation}
\mathbb{I}[x] = \begin{cases} 
1, & \text{if } x \text{ is true} \\
0, & \text{if } x \text{ is false}
\end{cases}
\end{equation}

In this paper, we place particular emphasis on the performance aspects of target classification. Therefore, the goal of the PE-MMSC system is to maximize the accuracy of target classification while ensuring the quality of data reconstruction via optimize the codec parameters. The system objective is:
\begin{subequations}
\begin{align}
& \max_{\psi, \varphi,\omega} \text{A}(\boldsymbol{C}_N^{\text{fin}},\boldsymbol{C}_N^{\text{true}}) \label{eq:objective} \\
 \text{s.t.} \quad &\text{NMSE}(\boldsymbol{D}_N^{\text{mod}},\boldsymbol{\hat{D}}_N^{\text{mod}} ) \leq \beta, \: \forall \text{mod}\label{eq:constraint}
\end{align}
\end{subequations}
where 
, with $\beta$ represents the The threshold of NMSE for data reconstruction. Owing to the complex data manifold of the sensing data, we propose employing a neural network (NN) based model to optimize the parameters and implement the functions.

\section{PE-MMSC Design}
In this section, a detailed design of the PE-MMSC is presented. Specifically, we focus on the fusion of multimodal semantics by the guidance of PE module with attention mechanism. 

\subsection{Perception Enhanced Module}
\label{pe:design}
\begin{figure}[ht]
    \centering
    \includegraphics[width=0.26\textwidth]{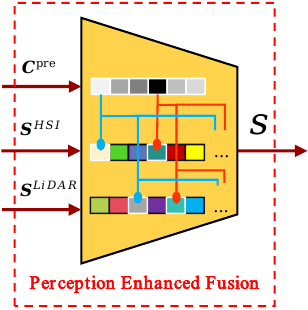}
    \caption{Attention mechanism of perception enhanced fusion}
    \label{fig:atten}
\end{figure}

The PE module is introduced in this section. In contrast to traditional semantic fusion approaches, where $\boldsymbol{S}^{\text{HSI}}$ and $\boldsymbol{S}^{\text{HSI}}$ are merely concatenated and input into the fusion module, we propose to input $\boldsymbol{C}^{\text{pre}}$ into the semantic fusion encoder. This serves as a guidance for the attention weight adjustment of each modality, as depicted in Fig.~\ref{fig:atten}, the intensity of $\boldsymbol{C}^{\text{pre}}$ reflects the prediction probabilities for each category, with red lines indicating enhanced semantic feature weights and blue lines indicating reduced semantic feature weights.
The attention mechanism of it, is $\boldsymbol{C}^{\text{pre}}$ provides an unbiased coarse classification probability distribution during the semantic fusion stage, $\boldsymbol{C}^{\text{pre}}$  plays a role in stage-wise supervision, effectively guiding the allocation of feature weights in the fusion process. During continuous training stage, the NN identifies semantic elements correlated with the max classification  probability $p_{\text{max}}^{\text{pre}}$ and strengthens their weights through binding mechanisms. Conversely, semantic elements related to min classification  probability $p_{\text{min}}^{\text{pre}}$ undergo weight reduction. This dynamic adjustment mechanism optimizes attention weights, ensuring that features strongly correlated with the primary classification obtain increased focus, thereby improving overall classification performance. 

$\boldsymbol{C}^{\text{pre}}$ provides the fusion guidance that is lacking in traditional multimodal fusion. This enables the fusion process to not solely rely on the self-adjustment of NNs, but rather to purposefully assign greater weights to more important semantic features based on the preliminary coarse classification result $[p_1^\text{pre},...,p_m^\text{pre}]$. Consequently, under the same SNR, the proportion of important information is increased, which enhances the robustness against noise and channel fading, thereby improving the overall task performance.

In summary, the role of PE-module in the semantic fusion process can be outlined in the following aspects:
\begin{itemize}
    \item[\textbullet] Guiding feature selection: $\boldsymbol{C}^{\text{pre}}$ enables the semantic fusion encoder to selectively strengthen the most relevant features to the task, while suppressing irrelevant features.
    \item[\textbullet] Improving fusion efficiency: by visualizing the weight changes of $\boldsymbol{C}^{\text{pre}}$ during the fusion process, researchers can gain a more intuitive understanding of the modality fusion between $\boldsymbol{S}^{\text{HSI}}$ and $\boldsymbol{S}^{\text{LiDAR}}$, which enhance the effectiveness of feature fusion. 
\end{itemize}

\subsection{Semantic Codec}
The NN architectures of the various semantic codecs are introduced in this section, including the semantic encoder, classifier, fusion encoder, and decoder, as well as the design of their respective loss functions.

\subsubsection{Semantic Encoder}
to facilitate $\boldsymbol{D}^{\text{mod}}$'s extension to additional modalities, an encoder based on the ReLU Block (RB) is utilized to extract $\boldsymbol{S}^{\text{mod}}$. Each RB comprises an FC layer for feature extraction, a BN layer to mitigate gradient vanishing and explosion, and a ReLU activation function to learn non-linear features from the data. Through multiple RB structures, the semantic encoder can gradually extract semantic features at varying levels. 
\subsubsection{Classifier}
to alieviate the computational of UAV, a single FC layer is employed to perform$f_{\psi^\text{pre}}^{\text{C}}(\cdot)$. For consistency in the PE-MMSC, the same layer is adopted for $f_{\psi^\text{fin}}^{\text{C}}(\cdot)$ at BS. The loss function of target classification module is cross entropy function, given by
\begin{equation}
L_{\psi,\varphi,\omega}(\boldsymbol{C}_N^{\text{phase}},\boldsymbol{C}_N^{\text{true}})= -\frac{1}{N} \sum_{i=1}^{N}\sum_{j=1}^{m}C_{i,j}^{\text{true}}log(C_{i,j}^{\text{phase}})
\end{equation}
where \( C_{i,j}^{\text{true}} \) and \( C_{i,j}^{\text{phase}} \) are the elements at the \( i \)-th row and \( j \)-th column of \( \boldsymbol{C}_N^{\text{true}} \) and \( \boldsymbol{C}_N^{\text{phase}} \).
\subsubsection{Semantic Fusion Encoder}
the structure of the semantic fusion encoder is similar to that of the encoder, but it requires fewer RBs because its primary function is to perform the fusion of semantic features rather than their extraction. The semantic fusion encoder fuses the complementary characteristics and semantic redundancy between different modalities, and attention mechanism of PE is employed to enhance the fusion process. 
\subsubsection{Semantic Decoder}
the decoder is constructed based on the Sigmoid Block (SB) and is designed to reconstruct $\boldsymbol{D}^{\text{mod}}$. It similarly incorporates an FC and BN layer; however, the activation function is replaced by a sigmoid function. This modification is because the Sigmoid function confines the output within the range of 0 and 1, which is beneficial for constraining the output range in data reconstruction tasks. For data reconstruction, the mean squared error (MSE) loss function is employed, gievn by
\begin{equation}
L_{\psi,\varphi,\omega}(\boldsymbol{D}_N^{\text{mod}},\boldsymbol{\hat{D}}_N^{\text{mod}})= \left\| \boldsymbol{D}_N^{\text{mod}} - \boldsymbol{\hat{D}}_N^{\text{mod}} \right\|^2
\end{equation}

We modify the traditional multimodal semantic fusion framework by changing the NN layer in \cite{13}, adding a PE module, and wireless channel training. A detailed description of each module's NN layers is provided in Table~\ref{tab:network}.
\begin{table}[h!]
\centering
\caption{Module NN description}
\label{tab:network} 
\begin{tabular}{|c|c|}
\hline
\textbf{Module}  & \textbf{Structure}                 \\ \hline
\textbf{RB block}              & FC + BatchNorm + ReLU             \\ \hline
\textbf{SB block}              & FC + BatchNorm + Sigmoid          \\ \hline
\textbf{Encoder}         & 4 RB block                 \\ \hline
\textbf{Decoder} & 2 SB block + FC                      \\ \hline
\textbf{Fusion Encoder}   & 2 RB  block                          \\ \hline
\textbf{Classifier}      & FC                                \\ \hline
\end{tabular}
\end{table}

\subsection{Training and Loss}
In conjunction with the optimization objectives (\ref{eq:objective}) and constraint (\ref{eq:constraint}), we formulate a joint training loss function, given by
\begin{align}
    & L_{\psi,\varphi,\omega}(\boldsymbol{C}_N^{\text{pre}}, \boldsymbol{C}_N^{\text{fin}}, \boldsymbol{D}_N^{\text{HSI}}, \boldsymbol{\hat{D}}_N^{\text{HSI}}, \boldsymbol{D}_N^{\text{LiDAR}}, \boldsymbol{\hat{D}}_N^{\text{LiDAR}}) \nonumber \\
    &  = \alpha_1 L_{\psi,\varphi,\omega}(\boldsymbol{C}_N^{\text{pre}},\boldsymbol{C}_N^{\text{true}}) \nonumber 
     + \alpha_2 L_{\psi,\varphi,\omega}(\boldsymbol{C}_N^{\text{fin}},\boldsymbol{C}_N^{\text{true}}) \nonumber \\
    &  + \alpha_3 L_{\psi,\varphi,\omega}(\boldsymbol{D}^{\text{HSI}}_N, \boldsymbol{\hat{D}}^{\text{HSI}}_N) 
     + \alpha_4 L_{\psi,\varphi,\omega}(\boldsymbol{D}^{\text{LiDAR}}_N, \boldsymbol{\hat{D}}^{\text{LiDAR}}_N)
\end{align}
where $\alpha_i$ represents the weight of different tasks,  $0\leq\alpha_i\leq1 $.

We opt to jointly train two classifiers and two decoders with the wireless channel. This approach not only enhances the NN's capability to acquire classification information but also ensures that the semantic features ultimately transmitted are not only robust against fading, but also tailored for multi-task robustness. Algorithm~\ref{alg:PE-MMSC} outlines the training process, which encompasses parameter initialization, network construction, extraction and fusion of semantic information, transmission of semantic features through the channel, reception and processing of information, calculation of the loss function and parameter updates through backpropagation and gradient descent.
\begin{algorithm}[ht]
\caption{Training algorithm of PE-MMSC}
\label{alg:PE-MMSC}
\begin{algorithmic}[1]
\STATE \textbf{Initialization:} Deploy a neural network and set up channels. Initialize the model parameters of $\psi$, $\varphi$, $\omega$.
\FOR{iteration of epoch \textbf{do}}
    \FOR{each batch of training data \textbf{do}}
        \STATE \textbf{Multimodal Sensing}:\\
        \quad get data $\boldsymbol{D}_N^{\text{HSI}}$ and $\boldsymbol{D}_N^{\text{LiDAR}}$ from training batches.
        \STATE \textbf{Semantic Extraction}:\\
        \quad input $\boldsymbol{D}_N^{\text{HSI}}$, $\boldsymbol{D}_N^{\text{LiDAR}}$, output $\boldsymbol{S}^\text{HSI}_N$, $\boldsymbol{S}^\text{LiDAR}_N$.
        \STATE \textbf{Perception Ehanced}:\\
        \quad input $\boldsymbol{S}^\text{HSI}_N$, output $\boldsymbol{C}^\text{pre}_N$.
        \STATE \textbf{Semantic Fusion}:\\
        \quad input $\boldsymbol{C}^\text{pre}_N$, $\boldsymbol{S}^\text{HSI}_N$, $\boldsymbol{S}^\text{LiDAR}_N$, output $\boldsymbol{S}_N$ 
        \STATE Transmit $\boldsymbol{S}_N$ through channel and BS receive $\boldsymbol{\hat{S}}_N$.
        \STATE \textbf{Target Classification}:\\
        \quad input $\boldsymbol{\hat{S}_N}$, output $\boldsymbol{C}^\text{fin}_N$
        \STATE \textbf{Data Reconstruction}:\\
        \quad input $\boldsymbol{\hat{S}}_N$, output $\boldsymbol{\hat{D}}_N^{\text{HSI}}$, $\boldsymbol{\hat{D}}_N^{\text{LiDAR}}$
        \STATE Calculate 
        \[
        L_{\psi, \varphi, \omega}(\boldsymbol{C}^\text{pre}_N, \boldsymbol{C}^\text{fin}_N, 
        \boldsymbol{C}^\text{true}_N
        \boldsymbol{D}^\text{HSI}_N, \boldsymbol{\hat{D}}^\text{HSI}_N, \boldsymbol{D}^\text{LiDAR}_N, \boldsymbol{\hat{D}}^\text{LiDAR}_N)
        \]
        \STATE Backpropagation and gradient descent, update the parameters $\psi$, $\varphi$, $\omega$.
    \ENDFOR
\ENDFOR
\end{algorithmic}
\end{algorithm}

\section{Experiments and Analysis}
\label{sec:EA}
\subsection{Dataset Description}
This study used the Houston2013 dataset~\cite{13}, which is widely used for performance evaluation in the field of remote sensing and has high-level qualitative and quantitative analytical value. The dataset integrates HSI and LiDAR technologies, covering a spectral range from 364 to 1,046 nanometers. The HSI data comprises 144 bands, while the LiDAR data consists of 349 bands and 1,905 pixels. The samples in the data set are annotated into 15 categories, including various scenes such as forests, water bodies, railways, etc., with each sample representing a pixel point. 

\subsection{Experimental Setup}
The proposed method is compared with the following baseline approaches:
\begin{itemize}
    \item[\textbullet] Traditional Multimodal Semantic Fusion Algorithm: The EndNet network presented in ~\cite{13} is used as the reference. The proposed PE-MMSC has more a PE module than the baseline line.
    \item[\textbullet] Deep EndNet: To validate that the performance improvement of the proposed method is due to the attention adjustment enabled by the PE module, rather than a deeper network architecture, we compared it with an EndNet network with a deepened semantic fusion encoder.
    \item[\textbullet] Single-modal Object Classification Algorithm: To assess the generalization capability of the proposed perception-enhanced design, we also compared the performance of networks with and without the PE module in a single-modal classification task.
\end{itemize}

All simulation methods strictly follow the system framework illustrated in Figure~\ref{fig:sys}. Additionally, the NN layer structure used is consistent with the hierarchical configuration listed in Table~\ref{tab:network}. However, in the joint encoder section, the deep network structure introduces additional RB to enhance the feature representation capability.
For different simulation schemes, the required floating-point operations are computed according to the methodology described in \cite{15} and detailed in Table~\ref{tab:flop}.

\begin{table}[h]
    \centering
    \caption{The Flops of various simulation methods}
    \begin{tabular}{|>{\centering\arraybackslash}m{2.5cm}|c|c|c|c|c|}
        \hline
        \makecell{\textbf{Module and} \\ \textbf{Neural Network}} 
        & \textbf{HSI} 
        & \textbf{LiDAR} 
        & \textbf{PE} 
        & \textbf{Fusion} 
        & \textbf{Total} \\
        \hline
        PE-MMSC    & 13536 & 11568 & 1920  & 43264  & 70288  \\
        \hline
        EndNet     & 13536 & 11568 & \diagbox{}{} & 41344  & 66448  \\
        \hline
        DeepEndNet & 13536 & 11568 & \diagbox{}{} & 107136 & 132240 \\
        \hline
        Hsi+PE     & 13536 & \diagbox{}{} & 1920  & 26880  & 42336  \\
        \hline
        LiDAR+PE      & \diagbox{}{} & 11568 & 1920  & 26880  & 40368  \\
        \hline
        HSI        & 13536 & \diagbox{}{} & \diagbox{}{} & \diagbox{}{} & 13536  \\
        \hline
        LiDAR      & \diagbox{}{} & 11568 & \diagbox{}{} & \diagbox{}{} & 11568  \\
        \hline
    \end{tabular}
    \label{tab:flop}
\end{table}

During the training process, the Adam optimizer is used, with a learning rate of $0.001$, a batch size of $64$, and a regularization parameter set to $\alpha = [0.6, 1, 1, 1]$. A total of $600$ iterations are executed.
Regarding channel modeling, this study utilizes a Rayleigh fading channel with $K=12$ for simulation experiments. Furthermore, to ensure semantic consistency across all schemes, all configurations are set to output semantic symbols $K=64$.

\subsection{Performance Evaluation And Analysis}

\begin{figure}[ht]
    \centering

    \begin{minipage}{0.49\linewidth}
        \centering
        \includegraphics[width=\linewidth]{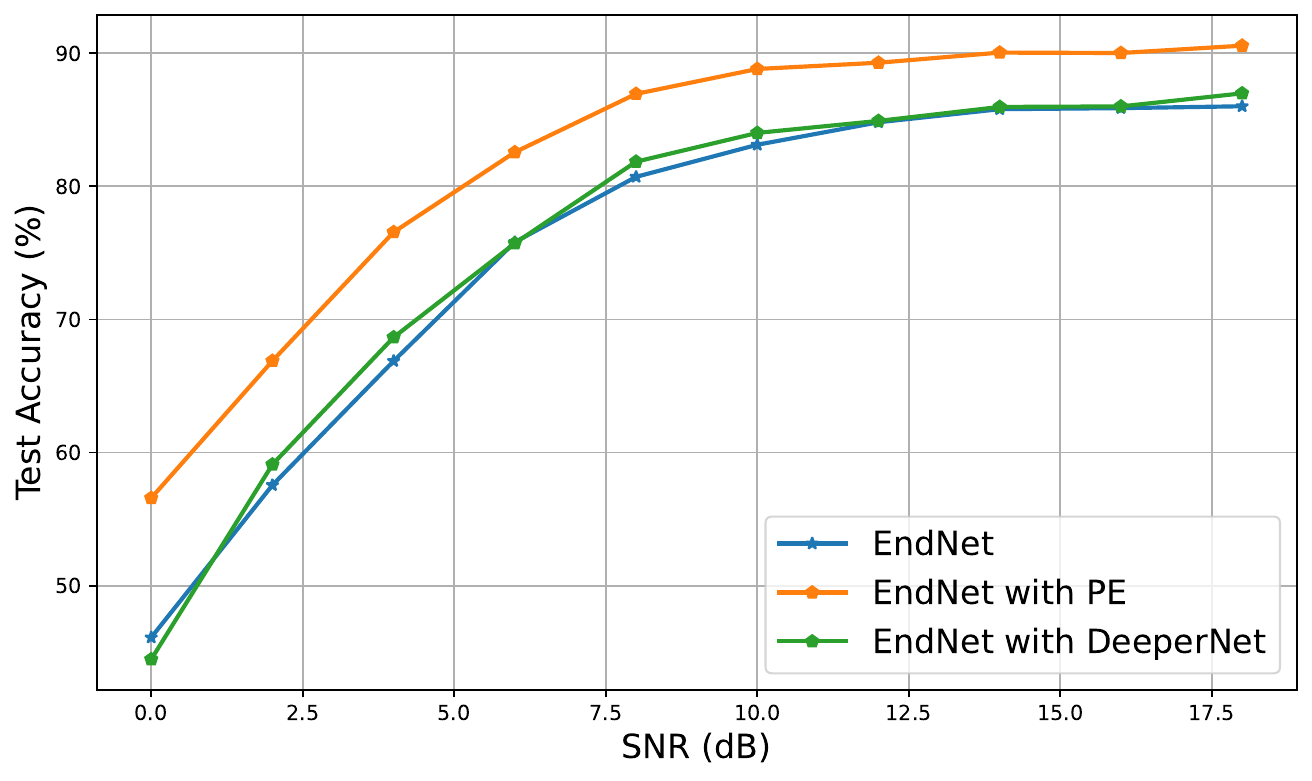}
        \captionsetup{justification=centering, font=small} 
        \caption{Accuracy vs SNR \newline (multi-modal)}
        \label{fig:image1}
    \end{minipage}
    \hfill
    \begin{minipage}{0.49\linewidth}
        \centering
        \includegraphics[width=\linewidth]{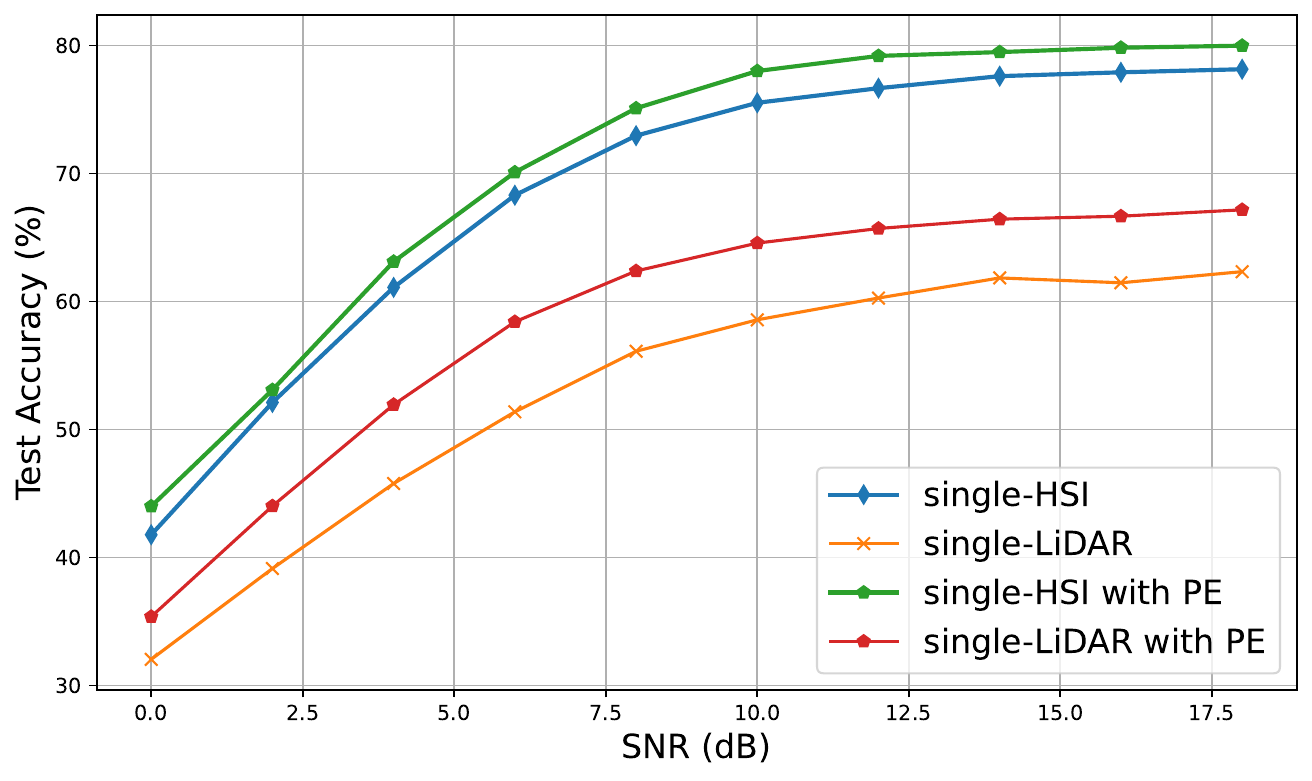}
        \captionsetup{justification=centering, font=small} 
        \caption{Accuracy vs SNR \newline (single-modal)}
        \label{fig:image2}
    \end{minipage}

    \vspace{5pt} 

    \begin{minipage}{0.49\linewidth}
        \centering
        \includegraphics[width=\linewidth]{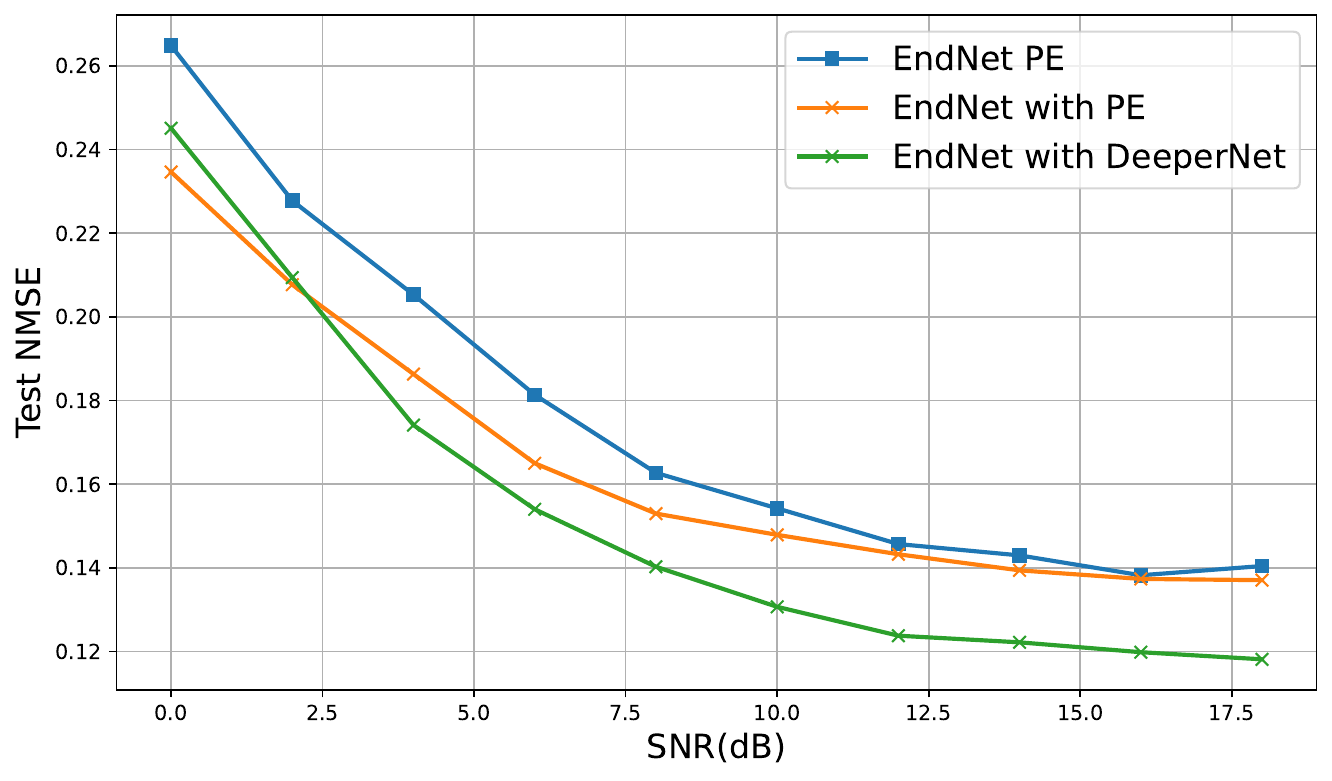}
        \captionsetup{justification=centering, font=small} 
        \caption{NMSE vs SNR (HSI)}
        \label{fig:image3}
    \end{minipage}
    \hfill
    \begin{minipage}{0.49\linewidth}
        \centering
        \includegraphics[width=\linewidth]{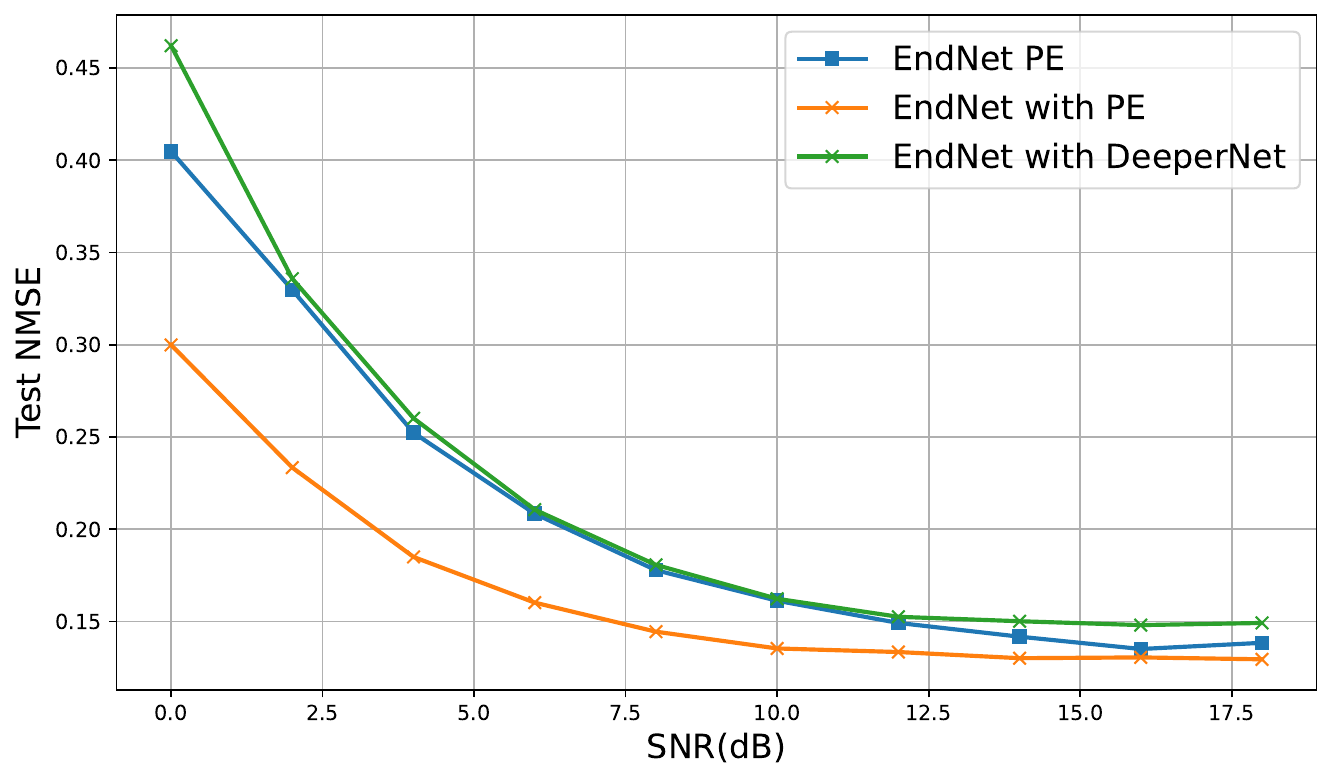}
        \captionsetup{justification=centering, font=small} 
        \caption{NMSE vs SNR (LiDAR)}
        \label{fig:image4}
    \end{minipage}

\end{figure}

Fig.~\ref{fig:image1}-\ref{fig:image4} illustrate the performance of various simulation methods in the target classification and data reconstruction task under Rayleigh fading channels with different SNR conditions. 

Fig.~\ref{fig:image1} shows that, the proposed PE-MMSC demonstrates a performance improvement of 5\% to 10\% compared to other methods across different SNR conditions. Notably, the performance enhancement is more significant under low SNR conditions, as the classification information retained in $\boldsymbol{C^\text{pre}}$ provides a more direct and valuable guidance, mitigating the impact of noise. Merely increasing the depth of NN, as seen in DeepEndNet, does not achieve the desired improvement

Fig.~\ref{fig:image2} further validates the generalizability of the proposed PE module. In classification tasks utilizing single-modal data, the PE module also exhibits significant performance improvements, with maximum gains of 3\% and 7\% for the HSI and LiDAR data, respectively.

Fig.~\ref{fig:image3} and~\ref{fig:image4} show the performance of PE-MMSC in data reconstruction tasks. Particularly for LiDAR data reconstruction, the proposed method achieves a maximum reduction in NMSE by 36\% compared to EndNet.

\begin{figure}[ht]
    \centering
    \includegraphics[width=0.4\textwidth]{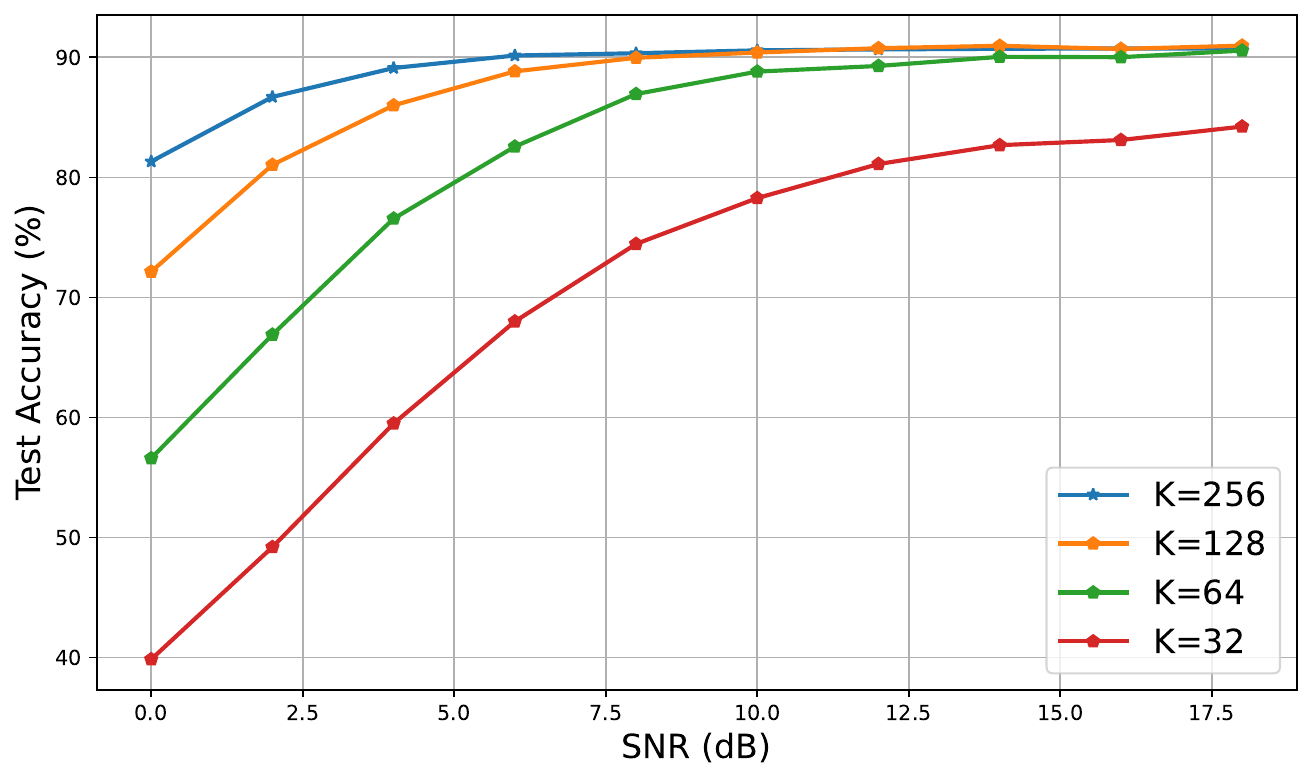}
    \caption{Different $K$ for Accuracy (PE-MMSC)}
    \label{fig:image5}
\end{figure}

We also evaluate the performance of the PE-MMSC system under different numbers of semantic symbols $K$, as illustrated in Fig.~\ref{fig:image5}. This suggests that $K=64$ is sufficient to represent semantic information. This reveals a larger $K$ values are required to improve noise resistance at low SNR levels, whereas smaller $K$ values can be adopted to optimize efficiency at high SNR levels.

A comprehensive analysis of Table~\ref{tab:flop} reveals that the proposed PE-MMSC method associated computational overhead is bearable, while the performance gain remains substantial. Table~\ref{tab:flop} shows underscores the effectiveness of the proposed method in balancing computational complexity and task performance.

\section{Conclusion}
In this paper, we investigates an efficient PE-MMSC system for ISAC UAVs. In our considered system, we extract and fusion the semantic features of sensing data. The use of the PE module effectively adjusts the attention during the multimodal semantic fusion process, thereby enhancing performance in multitask scenarios. Simulation results demonstrate that the proposed PE-MMSC significantly improves classification accuracy by 5\% to 10\% with only a minimal additional computational overhead, while ensuring data reconstruction quality. 
\section{Funding}
This work is supported in part by the National Natural Science Foundation of China under Grant 62271065 and U22B2001, and Beijing Natural Science Foundation under L242084.
\begingroup
\small 
\def\baselinestretch{1.}
\bibliographystyle{IEEEtran}
\bibliography{references}
\endgroup

\end{document}